\documentclass{ws-procs9x6}
\begin{document}

%
\def\cosech{\rm cosech}
\def\sech{\rm sech}
\def\coth{\rm coth}
\def\tanh{\rm tanh}
\def\half{{1\over 2}}
\def\third{{1\over3}}
\def\fourth{{1\over4}}
\def\fifth{{1\over5}}
\def\sixth{{1\over6}}
\def\seventh{{1\over7}}
\def\eigth{{1\over8}}
\def\ninth{{1\over9}}
\def\tenth{{1\over10}}
\def\bN{\mathop{\bf N}}
\def\R{{\rm I\!R}}
\def\Eins{{\mathchoice {\rm 1\mskip-4mu l} {\rm 1\mskip-4mu l}
{\rm 1\mskip-4.5mu l} {\rm 1\mskip-5mu l}}}
\def\Z{{\mathchoice {\hbox{$\sf\textstyle Z\kern-0.4em Z$}}
{\hbox{$\sf\textstyle Z\kern-0.4em Z$}}
{\hbox{$\sf\scriptstyle Z\kern-0.3em Z$}}
{\hbox{$\sf\scriptscriptstyle Z\kern-0.2em Z$}}}}
\def\abs#1{\left| #1\right|}
\def\com#1#2{
        \left[#1, #2\right]}
\def\square{\kern1pt\vbox{\hrule height 1.2pt\hbox{\vrule width 1.2pt
   \hskip 3pt\vbox{\vskip 6pt}\hskip 3pt\vrule width 0.6pt}
   \hrule height 0.6pt}\kern1pt}
      \def\boxop{{\raise-.25ex\hbox{\square}}}
\def\contract{\makebox[1.2em][c]{
        \mbox{\rule{.6em}{.01truein}\rule{.01truein}{.6em}}}}
\def\ltap{\ \raisebox{-.4ex}{\rlap{$\sim$}} \raisebox{.4ex}{$<$}\ }
\def\gtap{\ \raisebox{-.4ex}{\rlap{$\sim$}} \raisebox{.4ex}{$>$}\ }
\def\mn{{\mu\nu}}
\def\rs{{\rho\sigma}}
\newcommand{\Det}{{\rm Det}}
\def\Tr{{\rm Tr}\,}
\def\tr{{\rm tr}\,}
\def\sumij{\sum_{i<j}}
\def\e{\,{\rm e}}
\def\br{{\bf r}}
\def\bp{{\bf p}}
\def\bx{{\bf x}}
\def\by{{\bf y}}
\def\brhat{{\bf \hat r}}
\def\bv{{\bf v}}
\def\ba{{\bf a}}
\def\bE{{\bf E}}
\def\bB{{\bf B}}
\def\bA{{\bf A}}
\def\pa{\partial}
\def\dA{\partial^2}
\def\ddx{{d\over dx}}
\def\ddt{{d\over dt}}
\def\der#1#2{{d #1\over d#2}}
\def\lie{\hbox{\it \$}} 
\def\partder#1#2{{\partial #1\over\partial #2}}
\def\secder#1#2#3{{\partial^2 #1\over\partial #2 \partial #3}}
%
\def\be{\begin{equation}}
\def\ee{\end{equation}\noindent}
\def\bear{\begin{eqnarray}}
\def\ear{\end{eqnarray}\noindent}
\def\bec{\begin{equation}}
\def\eec{\end{equation}\noindent}
\def\bearc{\begin{eqnarray}}
\def\earc{\end{eqnarray}\noindent}
\def\benn{\begin{enumerate}}
\def\enn{\end{enumerate}}
\def\veject{\vfill\eject}
\def\ven{\vfill\eject\noindent}
%
\def\eq#1{{eq. (\ref{#1})}}
\def\eqs#1#2{{eqs. (\ref{#1}) -- (\ref{#2})}}
%
\def\totint{\int_{-\infty}^{\infty}}
\def\posint{\int_0^{\infty}}
\def\negint{\int_{-\infty}^0}
\def\pint{{\dps\int}{dp_i\over {(2\pi)}^d}}
%
\newcommand{\GeV}{\mbox{GeV}}
\def\FFdual{F\cdot\tilde F}
\def\bra#1{\langle #1 |}
\def\ket#1{| #1 \rangle}
\def\braket#1#2{\langle {#1} \mid {#2} \rangle}
\def\vev#1{\langle #1 \rangle}
\def\rightvac{\mid 0\rangle}
\def\leftvac{\langle 0\mid}
\def\ihbar{{i\over\hbar}}
\def\ge{\hbox{$\gamma_1$}}
\def\gz{\hbox{$\gamma_2$}}
\def\gd{\hbox{$\gamma_3$}}
\def\go{\hbox{$\gamma_1$}}
\def\gt{\hbox{\$\gamma_2$}}
\def\gth{\hbox{$\gamma_3$}} 
\def\gf{\hbox{$\gamma_5\;$}}
\def\slash#1{#1\!\!\!\raise.15ex\hbox {/}}
\newcommand{\slD}{\,\raise.15ex\hbox{$/$}\kern-.27em\hbox{$\!\!\!D$}}
\newcommand{\slpartial}{\raise.15ex\hbox{$/$}\kern-.57em\hbox{$\partial$}}
\newcommand{\PP}{\cal P}
\newcommand{\G}{{\cal G}}
\newcommand{\nc}{\newcommand}
\nc{\spa}[3]{\left\langle#1\,#3\right\rangle}
\nc{\spb}[3]{\left[#1\,#3\right]}
\nc{\ksl}{\not{\hbox{\kern-2.3pt $k$}}}
\nc{\hf}{\textstyle{1\over2}}
\nc{\pol}{\varepsilon}
\nc{\tq}{{\tilde q}}
\nc{\esl}{\not{\hbox{\kern-2.3pt $\pol$}}}
\newcommand{\cL}{\cal L}
\newcommand{\D}{\cal D}
\newcommand{\Dhalf}{{D\over 2}}
\def\eps{\epsilon}
\def\epshalf{{\epsilon\over 2}}
\def\lag{( -\partial^2 + V)}
\def\freeexp{{\rm e}^{-\int_0^Td\tau {1\over 4}\dot x^2}}
\def\kinb{{1\over 4}\dot x^2}
\def\kinf{{1\over 2}\psi\dot\psi}
\def\expk{{\rm exp}\biggl[\,\sum_{i<j=1}^4 G_{Bij}k_i\cdot k_j\biggr]}
\def\expp{{\rm exp}\biggl[\,\sum_{i<j=1}^4 G_{Bij}p_i\cdot p_j\biggr]}
\def\expshort{{\e}^{\half G_{Bij}k_i\cdot k_j}}
\def\expabb{{\e}^{(\cdot )}}
\def\epseps#1#2{\varepsilon_{#1}\cdot \varepsilon_{#2}}
\def\epsk#1#2{\varepsilon_{#1}\cdot k_{#2}}
\def\kk#1#2{k_{#1}\cdot k_{#2}}
\def\G#1#2{G_{B#1#2}}
\def\Gp#1#2{{\dot G_{B#1#2}}}
\def\GF#1#2{G_{F#1#2}}
\def\Dab{{(x_a-x_b)}}
\def\Dsq{{({(x_a-x_b)}^2)}}
\def\PITD{{(4\pi T)}^{-{D\over 2}}}
\def\4piTD{{(4\pi T)}^{-{D\over 2}}}
\def\4piT4{{(4\pi T)}^{-2}}
\def\TintmD{{\dps\int_{0}^{\infty}}{dT\over T}\,e^{-m^2T}
    {(4\pi T)}^{-{D\over 2}}}
\def\Tintm4{{\dps\int_{0}^{\infty}}{dT\over T}\,e^{-m^2T}
    {(4\pi T)}^{-2}}
\def\Tintm{{\dps\int_{0}^{\infty}}{dT\over T}\,e^{-m^2T}}
\def\Tint{{\dps\int_{0}^{\infty}}{dT\over T}}
\def\np{n_{+}}
\def\nm{n_{-}}
\def\Np{N_{+}}
\def\Nm{N_{-}}
\newcommand{\slG}{{{\dot G}\!\!\!\! \raise.15ex\hbox {/}}}
\newcommand{\Gd}{{\dot G}}
\newcommand{\Gund}{{\underline{\dot G}}}
\newcommand{\Gdd}{{\ddot G}}
\def\GBd12{{\dot G}_{B12}}
\def\Dx{\dps\int{\cal D}x}
\def\Dy{\dps\int{\cal D}y}
\def\Dpsi{\dps\int{\cal D}\psi}
\def\dint#1{\int\!\!\!\!\!\int\limits_{\!\!#1}}
\def\ddtau{{d\over d\tau}}
\def\ie{\hbox{$\textstyle{\int_1}$}}
\def\iz{\hbox{$\textstyle{\int_2}$}}
\def\id{\hbox{$\textstyle{\int_3}$}}
\def\ldop{\hbox{$\lbrace\mskip -4.5mu\mid$}}
\def\rdop{\hbox{$\mid\mskip -4.3mu\rbrace$}}
%
\newcommand{\1}{{\'\i}}
\newcommand{\no}{\noindent}
\def\non{\nonumber}
\def\cosech{\rm cosech}
\def\sech{\rm sech}
\def\coth{\rm coth}
\def\tanh{\rm tanh}
\def\half{{1\over 2}}
\def\third{{1\over3}}
\def\fourth{{1\over4}}
\def\fifth{{1\over5}}
\def\sixth{{1\over6}}
\def\seventh{{1\over7}}
\def\eigth{{1\over8}}
\def\ninth{{1\over9}}
\def\tenth{{1\over10}}
\def\bN{\mathop{\bf N}}
\def\R{{\rm I\!R}}
\def\Eins{{\mathchoice {\rm 1\mskip-4mu l} {\rm 1\mskip-4mu l}
{\rm 1\mskip-4.5mu l} {\rm 1\mskip-5mu l}}}
\def\Z{{\mathchoice {\hbox{$\sf\textstyle Z\kern-0.4em Z$}}
{\hbox{$\sf\textstyle Z\kern-0.4em Z$}}
{\hbox{$\sf\scriptstyle Z\kern-0.3em Z$}}
{\hbox{$\sf\scriptscriptstyle Z\kern-0.2em Z$}}}}
\def\abs#1{\left| #1\right|}
\def\com#1#2{
        \left[#1, #2\right]}
\def\square{\kern1pt\vbox{\hrule height 1.2pt\hbox{\vrule width 1.2pt
   \hskip 3pt\vbox{\vskip 6pt}\hskip 3pt\vrule width 0.6pt}
   \hrule height 0.6pt}\kern1pt}
      \def\boxop{{\raise-.25ex\hbox{\square}}}
\def\contract{\makebox[1.2em][c]{
        \mbox{\rule{.6em}{.01truein}\rule{.01truein}{.6em}}}}
\def\ltap{\ \raisebox{-.4ex}{\rlap{$\sim$}} \raisebox{.4ex}{$<$}\ }
\def\gtap{\ \raisebox{-.4ex}{\rlap{$\sim$}} \raisebox{.4ex}{$>$}\ }
\def\mn{{\mu\nu}}
\def\rs{{\rho\sigma}}
\def\Tr{{\rm Tr}\,}
\def\tr{{\rm tr}\,}
\def\sumij{\sum_{i<j}}
\def\e{\,{\rm e}}
\def\be{\begin{equation}}
\def\ee{\end{equation}\noindent}
\def\bear{\begin{eqnarray}}
\def\ear{\end{eqnarray}\noindent}
\def\bec{\begin{equation}}
\def\eec{\end{equation}\noindent}
\def\bearc{\begin{eqnarray}}
\def\earc{\end{eqnarray}\noindent}
\def\benn{\begin{enumerate}}
\def\enn{\end{enumerate}}
\def\ven{\vfill\eject\noindent}
%
\def\eq#1{{eq. (\ref{#1})}}
\def\eqs#1#2{{eqs. (\ref{#1}) -- (\ref{#2})}}
%
\def\totint{\int_{-\infty}^{\infty}}
\def\posint{\int_0^{\infty}}
\def\negint{\int_{-\infty}^0}
\def\pint{{\dps\int}{dp_i\over {(2\pi)}^d}}
%
\def\non{\nonumber}

\centerline{\Large\bf Three-loop Euler-Heisenberg Lagrangian and}
\vspace{2pt}
\centerline{\Large\bf asymptotic analysis in 1+1 QED}

\bigskip

\centerline{I. Huet$^a$, D.G.C. McKeon$^b$ and \underline{C. Schubert}$^a$}

\medskip

\begin{itemize}
\item [$^a$]
{\it 
Instituto de F\'{\i}sica y Matem\'aticas
\\
Universidad Michoacana de San Nicol\'as de Hidalgo\\
Edificio C-3, Apdo. Postal 2-82\\
C.P. 58040, Morelia, Michoac\'an, M\'exico\\
}
\item[$^b$]
{\it
Department of Applied Mathematics,\\
The University of Western Ontario,\\
London, ON N6A 5B7, Canada \\
}
\end{itemize}

\bigskip

\centerline{\sl Talk given by C. Schubert at QFEXT09, September 21 - 25, 2009}
\centerline{\sl (to appear in the conference proceedings)}

\bigskip

\noindent
{\bf Abstract:} In recent years, the Euler-Heisenberg Lagrangian has been shown
to be a useful tool for the analysis of the asymptotic growth of the
N-photon amplitudes at large N. Moreover, certain results and
conjectures on its imaginary part allow one,
using Borel analysis, to make predictions for those amplitudes
at large loop orders. Extending work by G.V. Dunne and one
of the authors to the three-loop level, but in the simpler context
of 1+1 dimensional QED, we calculate the corresponding Euler-Heisenberg
Lagrangian, analyse its weak field expansion, and study the congruence
with predictions obtained from worldline instantons. 
We discuss the relevance of these issues for Cvitanovic's conjecture. 


\vfill\eject\noindent

\bodymatter

\section{Cvitanovic's conjecture for g-2 in QED}

In their pioneering calculation of the $g-2$ factor of the electron to sixth
order in 1974, Cvitanovic and Kinoshita \cite{cvikin74} found a coefficient which was much
smaller numerically than had been expected by a naive estimate based on
the number of Feynman diagrams involved. A detailed analysis revealed 
extensive cancellations inside gauge invariant classes of diagrams. This led
Cvitanovic \cite{cvitanovic1977} to conjecture that, at least in the quenched
approximation (i.e. excluding diagrams involving virtual fermions) these cancellations
would be important enough numerically to render this series convergent
for the $g-2$ factor. 
Although nowadays there exist a multitude of good arguments against convergence of
the QED perturbation series (see, e.g., Ref. ~\refcite{dunnerev}), all of them are based
on the presence of an unlimited number of virtual fermions, so that Cvitanovic's conjecture
is still open today. Moreover, should it  hold true for the case of the $g-2$ factor,
it is natural to assume that it extends to arbitrary QED amplitudes in this quenched approximation.
In previous work \cite{dunschSD,colima} the QED effective
Lagrangian in a constant field was used for analyzing  the
$N$ -- photon amplitudes in the low-energy limit. Based on
existing high--order estimates for the imaginary part of this Lagrangian,
Borel dispersion relations, and a number of two--loop consistency checks, 
this very different line of reasoning makes ``quenched convergence'' appear
quite plausible for the case of the $N$ -- photon amplitudes. 
Its central point is an all-order conjecture for the imaginary
part of the constant-field effective Lagrangian for Scalar QED in the weak field 
limit due to Affleck, Alvarez, and Manton \cite{afalma} (AAM).
Here we present ongoing work towards a first three-loop check of this
conjecture \cite{wip}.

\section{The AAM conjecture}

Let us start with recalling the representation obtained by Euler and Heisenberg \cite{eulhei}
for the one-loop QED effective Lagrangian in a constant field,

\bear
{\cal L}^{(1)}_{\rm spin}(F) &=& - \frac{1}{8\pi^2}
\int_0^{\infty}{dT\over T^3}
\,\e^{-m^2T} 
\biggl[
{(eaT)(ebT)\over {\rm tanh} (eaT){\rm tan} (ebT)} 
- {1\over 3}(a^2-b^2)T^2 -1
\biggr]
\nonumber\\
\label{eulhei}
\ear
Here $T$ is the proper-time of the loop particle and $a,b$  are defined by 
 $a^2-b^2 = B^2-E^2,\quad  ab = {\bf E}\cdot {\bf B}$.
The analogous formula for Scalar QED was obtained by Weisskopf \cite{weisskopf}
but will also be called ``Euler-Heisenberg Lagrangian'' (EHL) in the following. 
Except for the magnetic case,
these effective Lagrangians have an imaginary part. 
Schwinger \cite{schwinger} found the following representation for the
imaginary parts in the purely electric case,

\begin{eqnarray}
{\rm Im} {\cal L}_{\rm spin}^{(1)}(E) &=&  \frac{m^4}{8\pi^3}
\beta^2\, \sum_{k=1}^\infty \frac{1}{k^2}
\,\exp\left[-\frac{\pi k}{\beta}\right] \non\\
{\rm Im}{\cal L}_{\rm scal}^{(1)}(E) 
&=&
-\frac{m^4}{16\pi^3}
\beta^2\, \sum_{k=1}^\infty \frac{(-1)^{k}}{k^2}
\,\exp\left[-\frac{\pi k}{\beta}\right]\non\\
\label{schwinger}
\end{eqnarray}
($\beta = eE/m^2$). These formulas imply that any constant electric field 
will lead to a certain probability for electron-positron pair creation from
vacuum. The inverse exponential dependence on the field
suggests to think of this as a tunneling process in which virtual pairs
draw enough energy from the field to turn real.
In the following we will be interested only in the
weak field limit $\beta \ll 1$, which allows us to truncate the 
series in (\ref{schwinger})
to the then dominant first ``Schwinger exponential''.

For the Scalar QED case,  Affleck et al. \cite{afalma} proposed in 1982 the
following all-loop generalization of (\ref{schwinger}),

\bear
{\rm Im}{\cal L}^{({\rm all-loop})}_{\rm scal}(E)
\,\,
&{\stackrel{\beta\to 0}{\sim}}&
\,\,
 \frac{m^4\beta^2}{16\pi^3}
\,{\rm exp}\Bigl[ -{\pi\over\beta}+\alpha\pi \Bigr] 
\label{ImLallloop}
\ear
This formula
is highly remarkable for various reasons.
Despite of its simplicity it is a true all-loop result; the rhs 
receives contributions from an infinite set of 
Feynman diagrams of arbitrary loop order,
including also mass renormalization counterdiagrams.
Moreover, the derivation given in Ref. \refcite{afalma} is very
simple, if formal. Based on a stationary path approximation of
Feynman's worldline path integral representation \cite{feynman} of
${\cal L}_{\rm scal}(E)$, it actually uses only
a one-loop semiclassical trajectory, and arguments that this
trajectory remains valid in the presence of virtual photon
insertions.
 
 An independent derivation of (\ref{ImLallloop}), as well as extension to
 the spinor QED case, was given by Lebedev
 and Ritus \cite{lebrit} through the consideration of higher-order corrections
 to the pair creation energy in the vacuum tunneling picture. 
 At the two-loop level,  (\ref{ImLallloop}) has 
 also been verified by  a direct calculation of the EHL \cite{dunsch1} (for the
 spinor QED case), as well as been extended to the case of a 
 self-dual field \cite{dunschSD}.
 
\section{Connection between the AAM and Cvitanovic conjectures}

Writing the AAM formula (\ref{ImLallloop}) as

\bear
{\rm Im}{\cal L}^{({\rm all-loop})}_{\rm scal}(E)\,\,
=
\,\,\sum_{l=1}^{\infty}{\rm Im}{\cal L}^{(l)}_{\rm scal}(E)
\,\,
&{\stackrel{\beta\to 0}{\sim}}&
\,\,
{\rm Im}{\cal L}^{(1)}_{\rm scal}(E)\,\,{\rm e}^{\alpha\pi}
\label{LalltoL1}
\ear
it states that an all-loop summation has produced the convergent
factor $\e^{\alpha\pi}$, clearly an observation similar in vein to Cvitanovic's.
Moreover, at least at a formal level it is not 
difficult to transfer this loop summation factor from 
${\rm Im}\Gamma(E)$ to the QED  photon amplitudes \cite{dunschSD,colima} .
Consider the weak field expansion of the $l$-loop contribution to the electric
EHL:

\bear
{\cal L}^{(l)}(E) &=& \sum_{n=2}^{\infty} c^{(l)}(n) \Bigl(\frac{eE}{m^2}\Bigr)^{2n}
\label{wfe}
\ear
Using Borel dispersion relations, (\ref{ImLallloop}) can be shown \cite{dunsch1,dunschSD} 
to imply that, at any fixed loop order $l$, the weak field expansion coefficients 
have the same asymptotic growth,

\bear
&& c^{(l)}(n)\quad {\stackrel{n\to \infty}{\sim}} \quad c^{(l)}_{\infty}\, \pi^{-2n}\Gamma[2n - 2]
\label{asymp}
\ear
where the constant $c^{(l)}_{\infty}$  relates directly to the prefactor of the
corresponding leading Schwinger exponential in the weak field limit:

\bear
{\rm Im}{\cal L}^{(l)}(E)\,\, &{\stackrel{\beta\to 0}{\sim}} &\,\,  c^{(l)}_{\infty}\,\e^{-\frac{\pi}{\beta}}
\label{borel}
\ear
As is well-known, the $n$th term in the weak field expansion of the $l$ - loop EHL
carries information on the corresponding  $N=2n$   - photon amplitudes in the low energy limit. 
Let us assume that the asymptotic behaviour should not depend on the choice of photon polarizations
$\varepsilon_i$ (this is plausible and supported by two-loop results \cite{colima}).  
Since the kinematical
structure of the $N$ - photon amplitudes in this limit reduces to a prefactor which
is the same at any loop order
\cite{mavisc}, one can eliminate it by dividing the $l$ - loop amplitude by the one-loop one.
Expanding (\ref{ImLallloop}) in $\alpha$ and combining it with (\ref{borel}) and (\ref{asymp}) 
one then  arrives at a formula for the ratio of amplitudes in the limit of large photon number,

\bear
{\rm lim}_{N\to\infty}
\frac
{\Gamma^{(l)}[k_1,\varepsilon_1;\ldots ;k_N,\varepsilon_N]}
{\Gamma^{(1)} [k_1,\varepsilon_1;\ldots ;k_N,\varepsilon_N]}
&=&
\frac{({\alpha\pi})^{l-1}}{(l-1)!}
\label{ratio}
\ear
If we could now sum both sides over $l$ and interchange the sum and
limit, we could reconstruct the $\e^{\alpha\pi}$ factor, and conclude
that the perturbation series for the $N$ - photon amplitudes,
at least in this low energy limit, is perfectly convergent!
But this is too good to be true, since so far we have nowhere made a
distinction between quenched and unquenched contributions to
the photon amplitudes, and convergence of the whole
perturbation series can certainly be excluded.  
However, as was noted in Ref. \refcite{colima} this distinction comes in
naturally if one takes into account that in the path integral derivation of
(\ref{ImLallloop}) in Ref. \refcite{afalma} the rhs comes entirely
from the quenched sector; all non-quenched contributions are suppressed
in the weak field limit. And since (switching back to the usual Feynman diagram
picture) the importance of non-quenched diagrams is growing with
increasing loop order, it is natural to assume that their inclusion will slow
down the convergence towards the asymptotic limit with increasing $l$,
sufficiently to invalidate the above naive interchange of limits. 
On the other hand, there is no obvious reason to expect such a slowing
down of convergence inside the quenched sector, which led to the
prediction \cite{colima} that Cvitanovic's ``quenched convergence''
will indeed be found to hold true for the photon amplitudes.

As a further step in this line of reasoning,
one should now check that the convergence of 
(\ref{asymp}) does not show a slowing down when
going from two to three loops if one keeps only quenched diagrams.
However, a calculation of any three-loop EHL,
be it in Scalar or Spinor QED, for an electric or self-dual field,
poses an enormous computational challenge.

Now, in 2006 M. Krasnansky \cite{krasnansky} calculated the two-loop EHL
in 1+1 dimensional Scalar QED and found it, surprisingly, to have a structure
almost identical to the one of the corresponding self-dual EHL in the
four-dimensional case:

\bear
{\cal L}_{\rm scal}^{(2)(4D)}(\kappa)
&=&
\alpha \,{m^4\over (4\pi)^3}\frac{1}{\kappa^2}\left[
{3\over 2}\xi^2 
-\xi'\right],\quad 
\xi(\kappa):= -\kappa\Bigl(\psi(\kappa)-\ln(\kappa)+\frac{1}{2\kappa}\Bigr)
\non\\
{\cal L}_{\rm scal}^{(2)(2D)}(\kappa)
&=&
-\frac{e^2}{32\pi^2}\left[
\xi^2_{2D} 
-4\kappa \xi_{2D}'\right] ,\quad 
\xi_{2D}:= -\Bigl(\psi(\kappa+\half)-\ln (\kappa)\Bigr)\non\\
\label{compscal2D4D}
\ear
($\psi(x)=\Gamma^\prime(x)/\Gamma(x)$, $\kappa :=  m^2/(2ef)$, 
$f^2=\fourth F_{\mu\nu}F^{\mu\nu}$).
This led us to consider 2D QED as a toy model for studying the above asymptotic predictions. 

\section{Extension of the AAM conjecture to 1+1 QED}

Of course, this will make sense only if the AAM formula (\ref{ImLallloop}) can be extended to the
2D case. The worldline instanton approach of \cite{afalma} can be
extended to the 2D case straightforwardly \cite{wip}, yielding the
following analogue of (\ref{ImLallloop}):

\bear
{\rm Im}{\cal L}(E)
&\sim&
\e^{-\frac{m^2\pi}{eE} + \tilde\alpha \pi^2  \kappa^2}
\label{ImLallloop2D}
\ear
($\tilde\alpha := 2e^2/\pi m^2$). 
Note that, contrary to the 4D case,  the second term in the exponent also involves the
external field. This leads also to a somewhat more complicated form of the 
corresponding asymptotic limit statement:

\bear
 {{\rm lim}_{n\to\infty}} {c^{(l)}(n)\over c^{(1)}(n+l-1)} 
&=& {(\tilde\alpha\pi^2)^{l-1}\over (l-1)!}
\label{asymp2D}
\ear

\section{Three loop Euler-Heisenberg Lagrangian in 1+1 QED}

At the one and two-loop level, we have obtained the 
EHL in 2D Spinor QED explicitly in terms of the gamma and 
digamma functions \cite{wip}:

\bear
{\cal L}^{(1)}(\kappa) &=& 
-{m^2\over 4\pi} {1\over\kappa}
\Bigl[{\rm ln}\Gamma(\kappa) - \kappa(\ln \kappa -1) +
\half \ln \bigl({\kappa\over 2\pi}\bigr)\Bigr]
\label{L2D1}\\
{\cal L}^{(2)}(\kappa) &=& {m^2\over 4\pi}\frac{\tilde\alpha}{4}
\Bigl[ \tilde\psi(\kappa) + \kappa \tilde\psi'(\kappa)
+\ln(\lambda_0 m^2) + \gamma + 2 \Bigr]
\label{L2D2}
\ear
Here $\tilde\psi(\kappa):=-\xi(\kappa)/\kappa$, and
$\lambda_0$ is an IR cutoff for the photon propagator which becomes
necessary at two loops in 2D.
Curiously, in the 2D case the two-loop Spinor QED result (\ref{L2D2})
is simpler (just linear in the digamma function) than the corresponding
Scalar QED one (\ref{compscal2D4D}).
Using the well-known large - $x$ expansion of ${\rm ln}\Gamma(x)$
in terms of the Bernoulli numbers $B_n$ one 
can then easily verify that (\ref{asymp2D}) does indeed hold true for
$l=2$.

\begin{figure}
\begin{center}
\epsfig{file=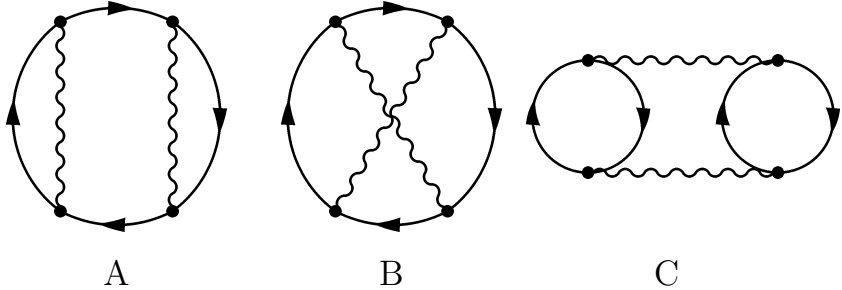,width=4.5in}
\end{center}
\caption{Diagrams contributing to the three-loop EHL}
\label{aba:fig1}
\end{figure}

At three loops our results are rather preliminary. 
There are three diagrams contributing to the
EHL, depicted in fig. 1 (the solid line represents the electron propagator
in a constant field). 
For all three we have obtained 
representation in terms of fourfold proper-time integrals.
The first six coefficients $c^{(3)}(n)$ for the quenched part
(diagrams A and B) were then
obtained in part analytically, in part by numerical integration. As it turns out,
at three loops all coefficients of the weak field expansion except the first one depend on the
IR cutoff $\lambda_0$. Introducing the modified cutoff 
$\Lambda := {\rm ln}(\lambda_0m^2)+\gamma$ the coefficients can
be written in the form 

\bear
c^{(3)}(n) &=& \tilde\alpha^2
\Bigl(
c^{(3)}_2 (n)\Lambda^2+ c^{(3)}_1(n)\Lambda + c^{(3)}_0(n)
\Bigr) 
\label{c3split}
\ear
where the coefficients $c^{(3)}_{1,2} (n)$ are rational
numbers, while $c^{(3)}_0(n)$ contains a $\zeta(3)$ already for $n=0$.
Since the prediction (\ref{asymp2D}) is cutoff-independent,
it can involve only the $c^{(3)}_0(n)$'s, so that the $c^{(3)}_{1,2}(n)$'s  
must be subdominant. For the series $c^{(3)}_2(n)$ we have been able to
compute a sufficient number of coefficients to verify that this is the case. Showing that the series
$c^{(3)}_0(n)$ indeed satisfies  (\ref{asymp2D}) is, however, not possible
with the coefficients obtained.

\section{Summary}

Extending the worldline instanton method of \cite{afalma} to 2D QED
we have obtained a prediction for the asymptotic growth of the weak field
expansion coefficients of the 2D EHL at any loop order.  At two loops
we have verified this prediction by an analytic calculation
of the EHL. At three loops we have obtained an integral representation
of the EHL suitable for a numerical calculation of the expansion coefficients,
and we expect to be able shortly to verify (or refute) the three main facts relevant for 
the AAM conjecture, namely  that (\ref{asymp2D}) holds at the $\tilde\alpha^2$ level,
independence of spin, and asymptotic suppression of the non-quenched diagram C.
On the slowing down issue, relevant for Cvitanovic's conjecture, it unfortunately
seems not to be possible to get information from the 2D QED case, due to the 
dependence of the three-loop expansion coefficients on the IR cutoff $\Lambda$;
although its numerical value does not affect the asymptotic limit, it
does have an influence on the rate of convergence towards it, which thus remains
ambiguous. Thus further progress in this line of attack on Cvitanovic's conjecture
presumably has to await the calculation of the three-loop EHL in 4D.


\begin{thebibliography}{99}
\bibitem{cvikin74}
P. Cvitanovic, T. Kinoshita, Phys. Rev. {\bf D 10} (1974) 4007.
\bibitem{cvitanovic1977}
P. Cvitanovic, Nucl. Phys. {\bf B 127} (1977) 176.
\bibitem{dunnerev}
G.V. Dunne, Cont. Adv. in QCD, {\bf 478} (2002) [hep-th/0207046].
\bibitem{dunschSD}
G.V. Dunne, C. Schubert, JHEP {\bf 0208} 
(2002) 053 [arXiv:hep-th/0205004]; JHEP {\bf 0206} (2002) 042 [arXiv:hep-th/0205005]. 
\bibitem{colima}
G.V. Dunne, C. Schubert, J. Phys.: Conf. Ser. {\bf 37} (2006) 59 [hep-th/0409021].
\bibitem{afalma}
I.K. Affleck, O. Alvarez, N.S. Manton, Nucl. Phys. {\bf B 197} (1982) 509.
\bibitem{wip}
I. Huet, D.G.C. McKeon, C. Schubert, in preparation.
\bibitem{eulhei}
W. Heisenberg, H. Euler, Z. Phys. {\bf 98} (1936) 714.
\bibitem{weisskopf}
V. Weisskopf, K. Dan. Vidensk. Selsk. Mat. Fy. Medd. {\bf 14} (1936) 1.
\bibitem{schwinger}
J. Schwinger, Phys. Rev. {\bf 82} (1951) 664.
\bibitem{feynman}
R.P. Feynman, Phys. Rev. {\bf 80} (1950) 440.
\bibitem{lebrit}
 S.L. Lebedev, V.I. Ritus,  Zh. Eksp. Teor. Fiz. {\bf 86} (1984) 408 
 [Sov. Phys. JETP 59 (1984) 237].
\bibitem{dunsch1}
G.V. Dunne, C. Schubert, Nucl. Phys. {\bf B 564} (2000) 591 [hep-th/9907190]. 
\bibitem{mavisc}
L.C. Martin, C. Schubert, V.M. Villanueva, 
Nucl. Phys. {\bf B 668} (2003) 335 [arXiv:hepth/0301022]. 
\bibitem{krasnansky}
M. Krasnansky,  Int. J. Mod. Phys. {\bf A 23} (2008) 5201 [hep-th/0607230].
\end{thebibliography}

\end{document}